\documentclass[preprint]{revtex4}
\usepackage{graphicx}

\newcommand{\bea}{\begin{eqnarray}}
\newcommand{\eea}{\end{eqnarray}}
\newcommand\beq{\begin{equation}}
\newcommand\eeq{\end{equation}}
\newcommand\beqa{\begin{eqnarray}}
\newcommand\eeqa{\end{eqnarray}}

\def\gtap{ \rlap{$>$}\lower5pt\hbox{$\sim$}}

\def\nc{N_c}

\newcommand{\ord}[1]{\ensuremath{\mathcal{O}\left(#1\right)}}

\newcommand{\mcal}[1]{\ensuremath{\mathcal{#1}}}
\def\OMIT#1{}

\begin{document}
\preprint{JLAB-THY-05-314}
\vspace{0.5cm}
\title{\phantom{x}
\vspace{0.5cm}  $\mathbf{1/N_c}$ Countings in Baryons: Mixings and Decays\footnote{Contribution to the \lq\lq Large $N_c$ QCD 2004\rq\rq proceedings.}}
\author{J. L. Goity }
\affiliation{Department of Physics, Hampton University, Hampton, VA 23668, USA. \\
 Thomas Jefferson National Accelerator Facility, Newport News, VA 23606, USA. }
\begin{abstract}
Based on a valence-quark picture of large $\nc$ baryons, I describe in some detail the $1/N_c$ power counting for   decays and spin-flavor configuration mixings in baryons.
\end{abstract}
\maketitle

\section{Introduction}
The application of the $1/\nc$ expansion of QCD  to phenomenology rests on our ability to determine the  $1/\nc$ power counting at the hadronic level.  In the  meson and glueball sector,  the power counting can be determined  by looking at the level of QCD Feynman diagrams where the  order in $1/\nc$ of each diagram is entirely determined by its topology \cite{tHooft}. The power counting is then translated to  a hadronic level  quantity by arguing that its  order in $1/\nc$ corresponds to that of the lowest order QCD Feynman diagrams that can contribute to that quantity (exception must be made in the chiral limit for quantities where there is a $\eta'$ pole contribution,  which gives an enhancement $\ord\nc$). In this way it is rather simple to setup the $1/\nc$ expansion in effective theories, such as in ChPT as described in several contributions to these proceedings \cite{ChPTtalks}. In the baryon sector the power counting is more involved  because it cannot be solely based on  topological arguments.  One  way to determine the power counting was proposed in  Witten's pioneering work on  baryons  \cite{Witten}, where a valence-quark picture of baryons is employed  with  the purpose of carrying out the combinatorics necessary to determine the power counting.   It should be noted that diagrams with quark loops are in general sub-leading, but they are not  necessarily  irrelevant in the large $\nc$ limit. For instance, while the mass of a baryon scales as $\nc$, the contributions by the quark sea are $\ord{\nc^0}$.   Also, there are   quantities where quark loops are crucial, such as the  strangeness form factors of the nucleon   where  a  strange-quark loop   gives  the dominant   contribution.
Establishing the $1/\nc$ power counting of such loop effects  can be easily achieved by  generalizing  the  valence-quark picture.

In this talk I addressed the $1/\nc$ power counting in the decays and configuration mixings where some novel results were recently obtained in Ref.~\cite{JLG2}, where more details can be found.
Let us start by giving a brief description of the non-relativistic valence quark picture of baryons.
In this  picture a baryon state is represented by
\beqa
\mid\Psi\;\rangle&=&\frac{1}{\nc!}\;\int \prod_{j=1}^{\nc}\; d^3 x_j\; \Psi_{\xi_1,\cdots,\xi_{\nc}}(x_1,\cdots,x_{\nc})\nonumber\\
&\times& \epsilon_{\alpha_1\cdots\alpha_{\nc}} \mid x_i \xi_1 \alpha_1;\cdots;x_{\nc} \xi_{\nc} \alpha_{\nc}\;\rangle,
\eeqa
with color indices $\alpha$ and spin-flavor indices $\xi$. The wave function $\Psi$ is totally symmetric under permutations of the indices $\{(x,\xi)\}$.
In the large $\nc$ limit  baryons are dense, and the valence picture can be implemented in the  Hartree approximation \cite{Witten} where the wave function is the product of single quark wave functions. For the ground state (GS) baryons the wave function will then read:
\beq
\Psi^{GS}(x,\xi)=\chi^S_{\xi_1,\cdots,\xi_{\nc}} \prod_{j=1}^{\nc}\;\phi(x_j),
\eeq
where $\chi^S$ is the totally symmetric spin-flavor wave function suitably normalized.
Excited baryons have one or more quarks in excited states. For baryons with a single excited quark the wave functions are:
\beqa
\Psi^{S}(x,\xi)&=& \frac{1}{\sqrt\nc }\;\chi^S_{\xi_1,\cdots,\xi_{\nc}}\;\sum_{i=1}^{\nc}        \phi(x_1)\cdots\phi'(x_i)\cdots\phi(x_{\nc}) \\
\Psi^{MS}(x,\xi)&=&\frac{1}{\sqrt\nc (\nc-1)!}\;\sum_{{\rm perm}~\sigma}\chi^{MS}_{\xi_{\sigma_1},\cdots,\xi_{\sigma_{\nc}}}\; \phi(x_{\sigma_1})\cdots\phi(x_{\sigma_{\nc-1}})\phi'(x_{\sigma_{\nc}}),\nonumber
\eeqa
where the two possible spin-flavor representations, namely the symmetric (S)  and the mixed-symmetric (MS)  representations are displayed. Note that the MS representation is the one  totally symmetric in the first $\nc-1$ indices.  Here the excited quark wave function $\phi'$ is taken to be orthogonal to $\phi$. These Hartree wave functions have the center of mass problem that can be handled by projecting them onto states of well defined total momentum. For instance, a Peierls-Thouless type  projection  adapts well to the large $\nc$ baryons.  Note that in the wave functions (2) and (3) the location of the center of mass has an uncertainty {\small{$\ord{1/\sqrt\nc}$}}. One then expects that  the error introduced by the CM problem will be sub-leading in $1/\nc$.   An analysis where this issue is addressed  in detail will be presented elsewhere. 

It is now straightforward to calculate operator matrix elements. For 1- and 2-body operators we have
the following master formulas \cite{JLG2}: 
\beqa
\langle \Psi'\mid \Gamma_1(x)\mid \Psi\rangle&=&N_c \int  \prod_{j=1}^{N_c-1} d^3 x_j \; \Psi'^*_{\xi_1,\cdots,\xi_{N_c-1},\xi'}(x_1,\cdots,x_{N_c-1},x)\nonumber\\
&\times& \Gamma_{\xi' \xi}(x)\;
\Psi_{\xi_1,\cdots,\xi_{N_c-1},\xi}(x_1,\cdots,x_{N_c-1},x)\nonumber\\
\langle \Psi'\mid \Gamma_2(x,y)\mid \Psi \rangle&=&
\frac{N_c-1}{N_c}
\int \prod_{j=1}^{N_c-2} d^3 x_j \\
&\times& \Psi'^*_{\xi_1,\cdots,\xi_{N_c-2},\xi'_{N_c-1},\xi'_{N_c}}
(x_1,\cdots,x_{N_c-2},x,y)\nonumber\\
&\times&\left(
\Gamma^{\xi'_{N_c} \alpha_{N_c-1},\xi'_{N_c-1} \alpha_{N_c}}_
{\xi_{N_c} \alpha_{N_c},\xi_{N_c-1} \alpha_{N_c-1}} (x,y)-\Gamma^{\xi'_{N_c} \alpha_{N_c},\xi'_{N_c-1} \alpha_{N_c-1}}_
{\xi_{N_c} \alpha_{N_c},\xi_{N_c-1} \alpha_{N_c-1}} (x,y)\right)\nonumber\\
&\times&\Psi_{\xi_1,\cdots,\xi_{N_c-2},\xi_{N_c-1},\xi_{N_c}}(x_1,\cdots,x_{N_c-2},x,y),\nonumber
\eeqa
where the 1- and 2-body operators are  represented by the color singlet  spin-flavor tensors $\Gamma_1$ and $\Gamma_2$. In the latter case two different contractions of the color indices
result, where one of them is in general sub-leading in $1/\nc$. 

For the sake of illustration, let us consider the  application  to the matrix elements of the axial current operator. The structure of the 1-body spin-flavor operator associated with the axial current is simply $g_{ia}\equiv\frac 14\;\sigma_i t_a$, where $\sigma_i$ and $t_a$ are respectively spin and flavor generators in  the fundamental representation. From  Eqn.~(4) one immediately obtains for matrix elements between GS baryons:
\beqa
\langle \Psi^{\rm GS'}\mid A_{ia}(x) \mid \Psi^{\rm GS}>&=& N_c \;\phi^*(x)\, \phi(x)\\
&\times&\chi'^{ S\dagger }_{\xi_1, \cdots, \xi_{N_c-1}, \xi'_{N_c}}\; (g_{ia})_{\xi'_{N_c}\xi_{N_c}}\;\chi^{ S }_{\xi_1, \cdots, \xi_{N_c-1}, \xi_{N_c}} .\nonumber
\eeqa
Using the total symmetry of the spin-flavor wave functions one has that 
\beq
\chi'^{ S\dagger }_{\xi_1, \cdots, \xi_{N_c-1}, \xi'_{N_c}}\; (g_{ia})_{\xi'_{N_c}\xi_{N_c}}\;\chi^{ S }_{\xi_1, \cdots, \xi_{N_c-1}, \xi_{N_c}}=\frac 1{\nc} \;\chi'^{ S^{\dagger} } \; G_{ia}\; \chi^{ S },
\eeq
where $G_{ia}$ is now the spin-flavor generator in the totally symmetric spin-flavor representation with $\nc$ indices.  The above matrix elements of  $G$  can be  $\ord{\nc}$ if the states have spin $\ord{1}$  ($G$ is said to be a coherent operator)\footnote{In $SU(4)$ all the matrix elements of $G$ are $\ord{\nc}$, while in $SU(6)$ they can also be $\ord{1}$ such as the matrix elements of $G_{i8}$ between non-strange baryons.  }.  From this follows the well known result  that the matrix elements of the axial currents between GS baryons carrying  spins of  $\ord{1}$ are $\ord{\nc}$.  This well known result is at the heart of the  Gervais-Sakita-Dashen-Manohar consistency relations \cite{ConsistencyRel} that imply the emergence of a dynamical contracted $SU(2 N_f)$ symmetry as $\nc\to\infty$. It should be emphasized that the valence quark picture automatically satisfies those consistency conditions. One implication of the dynamical symmetry is that the GS baryons must form a tower of degenerate states as $\nc\to\infty$.  In particular,   the mass splittings   between the lower lying spin states must be $\ord{1/\nc}$.  These splittings are primarily  produced by spin-spin (hyperfine) interactions  as the hyperfine term derived from the one-gluon exchange that is given by  the 2-body operator
\beqa
{\mathcal{H}}_{HF}(x-y)&=&-g^2\;\frac{1}{4 m_q^2}\left(\pi\;\delta^3(x-y)\;\delta_{ij}+\cdots\right)\;\sigma_i     \lambda^A \otimes  \sigma_j     \lambda^A ,
\eeqa 
where $\lambda^A$ are the color generators and the ellipsis denote further tensor terms. The important point here is that in order to have a 2-body operator one pays a the price of a factor $g^2=\ord{1/\nc}$. An explicit calculation using Eqn.~(4) shows that in fact the hyperfine interaction splits  states of low spin by amounts $\ord{1/\nc}$. 

Let us consider another application, this time to  excited baryons.  In excited baryons where one excited quark is in an $\ell>0$ state there is   spin-orbit interaction. It is of interest to determine the order at which it contributes. In the Hartree picture one can identify the spin-orbit  interaction due to the Thomas precession. This term is determined by  the effective  Hartree interaction, and because this interaction is $\ord{1}$   the spin-orbit term is also $\ord{1}$. The simplest term for the spin-orbit interaction that can written  is the 1-body operator:
\beq
H_{SO}=w(r)\;\vec{L} \cdot \vec\sigma.
\eeq
Because  the excited quark is the only one that can carry  orbital angular momentum,  $H_{SO}$ only affects  the excited quark.  Using  Eqn.~(4) for an excited state with one excited quark, one immediately realizes that the $1/\nc$ power  is determined by  the matrix elements
\beq
\langle \chi\mid \sigma^*_i\mid\chi\rangle\equiv \chi^\dagger_{\xi_1,\cdots,\xi_{\nc}} \sigma_{_{\xi_{\nc}\xi'_{\nc}}} \chi_{_{\xi_1,\cdots,\xi'_{\nc}} },
\eeq
where $\chi$ is the spin-flavor wave function of the excited baryon, and $\sigma^*$ indicates that the spin operator is acting only on the spin of the excited quark. It is straightforward to show that for a totally symmetric $\chi$ the result is $\ord{1/\nc}$, and for a mixed symmetric $\chi$ the result is $\ord{1}$. This implies that 
\beq
\langle \Psi\mid H_{SO}\mid \Psi\rangle =\left\{\begin{array}{l c l}
\ord{1\over\nc}&~~~~ \rm if ~\chi~\rm is & \rm S\\
\ord{1}&~~~~\rm  if~\chi~\rm is& \rm MS
\end{array}\right. 
\eeq
The importance of this result is that in excited baryons the spin-flavor symmetry present for the GS baryons can be broken at zeroth order.  Several works have studied the implications of the zeroth order breaking \cite{Masses}, which always involves  coupling to $\vec{L}$. The interesting conclusion drawn from analyzing excited baryon masses is that, for dynamical reasons, all mass operators involving orbital couplings are suppressed, the effects being smaller than the sub-leading hyperfine ones. This dynamical property of QCD is somewhat mysterious and represents an interesting open problem to be solved. It should be pointed out that the weakness of orbital couplings allows one to make use of an approximate $O(3)\times SU(2 N_f)$ symmetry in describing excited baryons, and thus assign excited baryons in multiplets of that symmetry.  It is in fact well known that the  established excited baryons   fit very well into such a multiplet structure. One very interesting point of consistency that can be made is that the spin-orbit splittings in S representation states should be suppressed by a factor $1/\nc$ with respect to similar splittings in MS states. This is clearly seen by comparing the observed spin-orbit splittings in the  states assigned to the $SU(6)$ 70-plet and the ones assigned to the 56-plet \cite{Masses}.  While these splittings are observed to be small in the 70-plet, they are actually almost insignificant in the known 56-plets.

\section {Decays}
The decays of excited baryons proceed primarily via the emission of  a meson. Let us analyze the transitions mediated by  a single $\pi$, $K$ or $\eta$ meson.  We consider here a picture in which these mesons couple to the valence quarks via a 1-body operator as it has been proposed in the chiral quark model \cite{ChQM}, namely
\beq
H_{\rm{ChQM}}=-\frac{g_A^q}{ F_\pi} \int d^3 x\;\partial_i\pi_a\;q^\dagger(x) g_{ia} q(x),
\eeq
where $g_A^q=\ord{1}$ is the quark axial coupling and $F_\pi=\ord{\sqrt{\nc}}$ is the pion decay constant.  From  Eqn.~(5) one concludes that the pseudo-scalar mesons have couplings $\ord{\sqrt\nc}$ to GS baryons (some of the couplings are {\small{$\ord{1/\sqrt\nc}$}} as in the case of the $\eta$ couplings to non-strange baryons). The reason the GS baryons are narrow in large $\nc$ limit is the phase space suppression factor $1/\nc^3$ that results for P-wave transitions, giving in the end a width order $1/\nc^2$.   Similar suppressions take place for transitions between states within  the same excited multiplet.

Let us now discuss transitions from excited baryons to GS baryons. From  Eqns.~(4) and (11) the decay amplitude is given by:
\beqa
\langle \Psi^{\rm GS}+\pi_a\mid\Psi'\,\rangle&=& \frac{g_A^q}{F_\pi} \;\sqrt\nc\; k_{\pi\stackrel{}{i}}\nonumber\\
&\times& \int d^x\;e^{i k_\pi \cdot x}\;\phi^*(x) \phi'(x)\;\langle \chi^S\mid g_{ia}\mid \chi'\rangle,
\eeqa
where
 \beq
 \langle \chi^S\mid g_{ia}\mid \chi'\rangle \equiv \chi^{S^*}_{\xi_1,\cdots,\xi'_{\nc} }(g_{ia})_{\xi'_{\nc} \xi_{\nc}} \chi'_{\xi_1,\cdots,\xi_{\nc}} 
 \eeq
  Here  the last index in $\chi'$ is the one associated with the excited quark. Irrespective of the representation $\chi'$, these spin-flavor matrix elements are $\ord{1}$, and thus the decay amplitude is also $\ord{1}$. This represents an important general conclusion: excited baryons, unlike excited mesons, are not narrow in large $\nc$.  This fact has profound significance as it implies that excited baryons can also be seen as resonances in GS baryon-meson scattering \cite{Lebed}. Analyses of the widths of the  negative parity  $SU(6)$ 70-plet baryons \cite{CaroneDecays,GSS4} and the positive parity 56-plet \cite{CarlsonCarone,SGDecays} indicate the dominance of the 1-body operator amplitude in these cases, although in the 70-plet case there is need for the $1/\nc$ corrections, that  require 2-body operators, in order to improve the fit to the D-wave partial widths. A recent analysis of the decays of $\ell=2$ positive parity baryon decays \cite{SGDecays}, however,  shows that  2-body operators are important for a consistent fit.  An interesting consistency test provided by the decays are the decays with emission of an $\eta$ meson, where non-strange states in the 70-plet have in general amplitudes $\ord{1}$  while the non-strange states in the  56-plet have amplitudes $\ord{1/\nc}$ \cite{SGDecays}. The suppression of these latter decays is  experimentally well established \cite{PDG}.\\~\\
  \begin{figure}[ht]
\begin{center}
\includegraphics[width=9cm,height=2cm]{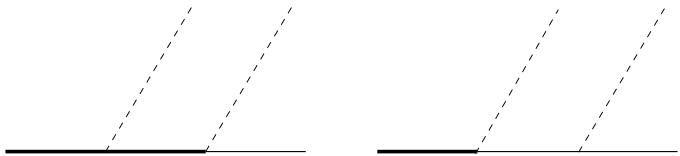}
\caption{\small The thick solid line represents  excited baryons belonging to a single multiplet, the thin one represents  a ground-state baryon, and the dashed lines represent pions. The vertices connecting an excited and ground-state baryon are proportional to  $1/\sqrt{N_c}$ for two-quark excited baryons,  while the other vertices are proportional to $\sqrt{N_c}$.}
\label{fig1} 
\end{center}
\end{figure}

One may wonder what happens with the emission of two pions in the form shown in Fig.~1. 
the individual diagrams are order $\sqrt{\nc}$, which implies that there must be a cancellation
of the terms of that order between the two diagrams. Indeed, such a cancellation was pointed out by Pirjol and Yan \cite{PirjolYan} and can be readily shown how it occurs in the valence picture:
\beqa
\langle \Psi^{GS}+\pi_a+\pi_b\mid \Psi'\;\rangle & \propto& \frac{\sqrt{\nc}}{F_\pi^2}\;
\left(\frac{k_i^i k_2^j}{k_1^0}\langle \chi\mid g_{jb}\mid \chi'_n\rangle\langle \chi'_n\mid G_{ia}\mid\chi'\rangle\right.\\
&-&\left. \langle \chi\mid G_{ia}\mid \chi_n\rangle\langle \chi_n\mid g_{jb}\mid\chi'\rangle+(1\leftrightarrow 2, a\leftrightarrow b)\phantom{\frac{\frac{}{}}{\frac{}{}}}\!\!\!\right)\nonumber
\eeqa
where $k_{1,2}$ are the pion momenta, and the sum over intermediate states is over states in the same multiplet  as the excited state (first term) and over GS (second term).   Explicit calculation shows that for an excited state belonging to the S representation the amplitude is $\ord{1/{\nc}^{3/2}}$, and $\ord{1/\sqrt{\nc}}$ if it belongs to the MS. Thus, this type of two-meson emission is subleading. It is however interesting to observe that there are decays where the two-pion channel  is important  (we exclude the two-pion channel of the chain type  such as $N^*\to\Delta\pi\to N\pi\pi$).  It seems clear that the two-pion decays are dominated by processes involving an intermediate meson, such as a $\rho$ meson, with the pions being the decay products of the intermediate meson.  Examples of states where  the two-pion channel is important are  $N^*(1520)$, $N^*(1700)$, $N^*(1720)$, $\Delta(1700)$, $\Delta(1710)$ and  $\Delta(1720)$ \cite{PDG}.  In all these  cases one can see the intermediate meson dominance mechanism at work.  Although no detailed analysis of the emission of $\rho$ mesons in the  $1/\nc$ expansion has been carried out, it is clear  that they are  $\ord{1}$.  This can be seen by using a simple model where the   $\rho$  couples  to quarks via   the 1-body operator  $1/F_\rho\ \epsilon_{ijk}\, q^\dagger\, g_{ia}\,q\, \partial_j \rho_{i}^a $, where $F_\rho$ is the decay constant.  It is straightforward to check that this operator leads to amplitudes $\ord{1}$ for  transitions from excited baryon to GS baryon and a $\rho$ meson.

Other transitions of interest where the $1/\nc$ expansion gives important insights are  the transitions between excited  multiplets. One immediately observes that there is a factor $1/\sqrt\nc$ in the matrix element for each excited quark in the baryons. Indeed, an explicit evaluation leads to the general form:
\beq
\langle\; \Psi'+\pi_a\mid \Psi^"\;\rangle=\frac{g_A^q}{F_\pi}\;k_{\pi i}\;\langle \chi'\mid g_{ia}\mid \chi"\rangle \;\int d^3x \; e^{ik_\pi\cdot x} \phi'^*(x)\phi"(x).
\eeq 
This amplitude is  $\ord{1/\sqrt{\nc}}$, meaning  that the partial widths for such excited-to-excited transitions are $\ord{1/\nc}$.  Thus, in  large $\nc$ limit excited baryons decay directly to GS baryons  while  cascade decays are  suppressed.  It is expect that this prediction based on the Hartree picture used here is correct for QCD.  However, one can immediately see  a puzzle emerging,  which may not be relevant for $\nc=3$, but it requires discussion in large $\nc$. If one calculates the transition amplitude for an excited baryon with two excited quarks to decay into a GS baryon and  a pion, the amplitude turns out to be  $\ord{1/\sqrt{\nc}}$. On the other hand, one expects   in general that excited baryons are not stable in large $\nc$. At this point it is not clear to us  what improvement over the Hartree picture used here  could solve this  puzzle.

\section{ Mixings}

In this section we discuss the problem of configuration mixings, namely mixings between different representations of $O(3)\times SU(2N_f)$. As mentioned earlier, the weakness of orbital couplings at $\ord{1}$ makes the classification of states in multiplets  of this group very convenient. In the strict large $\nc$ limit one should, however,  carry out the analysis in a different way, as it was pointed out in  \cite{PirjolSchat}.
Here we focus  on mixings involving GS baryons and excited baryons with only one excited quark.
The Hamiltonian that drives the mixings must be a scalar, so it transforms under $O(3)\times SU_{\rm spin}(2)$ as $(j, j)$. 

By simple inspection one finds that there is only one 1-body operator that can produce mixing, namely the spin-orbit operator of the general form shown in Eqn.~(8). This operator transforms as $(1,1)$ and  gives $\Delta\ell=0$ mixings. It only affects states with $\ell>0$, and thus it can only mix excited states.  Therefore, the only  relevant mixing that it can give is the S-MS spin flavor mixing. The typical matrix elements for the mixing can  be derived from Eqn.~(4) and are of the following form:
\beq
\langle MS, \ell \mid H_{SO}^{\rm mix} \mid S,\ell\rangle \propto \langle L_i \rangle \langle MS\mid s_i \mid S\rangle.
\eeq
For the sake of simplicity  we have disregarded coupling the spin and orbital angular momentum of the states to well defined total $J$, as this is unnecessary for our arguments and trivial to carry out.
The important point is now that the spin-flavor matrix element $ \langle MS\mid s_i \mid S\rangle$ is $\ord{1}$.  Thus, there is in principle $\Delta\ell=0$  configuration mixing at zeroth order in the $1/\nc$ expansion, as it was first pointed out in  \cite{CohenLebedNelore}.

The generic 2-body Hamiltonian that can contribute to mixing is 
\beq
{\mcal{H}}^{\rm mix}_{{\rm{2-body}}}(x,y)=\frac{1}{\nc}\; L_{ij}(x,y)\;{\mcal{S}}_{ij} ,
\eeq
where $ {\mcal{S}}$ is a spin-flavor tensor operator  which is a flavor singlet (we are disregarding any flavor symmetry breaking as it is not relevant for the discussion).
The first mixing of interest is the mixing between GS and excited baryons. Applying Eqn.~(4) one readily obtains
\beqa
\langle\Psi'\mid H^{\rm mix}_{\rm 2-body}\mid \Psi_{\rm GS}\rangle&=& \sqrt{\nc} \int d^3x d^3y (\phi^*(x)\phi'*(y)+x\leftrightarrow y)\nonumber\\
&\times& L_{ij}(x,y) \phi(x)\phi(y) \langle \chi'\mid {\mcal{S}}_{ij}\mid S\rangle,
\eeqa
where the 2-body spin-flavor matrix elements are specifically:
\beq
 \langle \chi'\mid {\mcal{S}}_{ij}\mid S\rangle \equiv \chi'_{\xi_1, \cdots,\xi'',\xi'''} {{\mcal{S}}_{ij}}^{\xi'' \xi''''}_{~\xi,\xi'} \chi^S_{\xi_1,\cdots,\xi,\xi'}.
\eeq
The order of the mixing amplitude in Eqn.~(18) is determined by the order of the spin-flavor matrix elements.
An explicit evaluation of the different 2-body tensors ${\mcal{S}}$ that can be built with the spin-flavor generators gives the results shown in  Table~1.
\begin{table}[h]
\caption{List of  2-body spin-flavor operators and matrix elements relevant to configuration mixings. Here $j$ indicates angular momentum of the operator.
  {\bf 1 } denotes the singlet spin-flavor operator. The asterisks  indicates entries that produce irrelevant configuration mixings.  The matrix elements are here defined in the way shown Eqn.~(19). }
{\setlength{\tabcolsep}{0.6em}
\begin{tabular}{@{}lll@{\hspace{0.12em}}c@{\hspace{0.12em}}l@{}}
\hline Operator & &\\
and~~~~ &{~~~~$j=0$}&{$~~~j=2$} \\ Matrix Element & & \\ \hline
$\langle S\mid s_i\otimes s_j\mid S\rangle$ & ~~~* $\ord{1/\nc} {\mathbf {1 }}+\ord{1/\nc^2}$  & ~~~$\ord{1/ N_c^2}$  \\
$\langle S\mid g_{ia} \otimes g_{ja}\mid S\rangle$ & ~~~* $\ord{ 1} {\mathbf {1 }}+\ord{ 1/N_c^2}$ & ~~~$\ord{ 1/N_c^2}$  \\
$\langle MS\mid   s_i\otimes s_j \mid S\rangle $ &~~~~ $\ord{ 1/N_c} $  & ~~~$\ord{ 1/N_c}$  \\
$\langle MS\mid g_{ia}\otimes g_{ja}\mid S\rangle $ &~~~~ $\ord{ 1/N_c}$& ~~~$\ord{ 1/N_c}$  \\
$\langle MS\mid s_i\otimes s_j\mid MS\rangle $ & ~~~* $\ord{1/N_c} \mathbf{1}+\ord{1/\nc^2} $   & ~~~$\ord{1/N_c}$  \\
$\langle MS\mid g_{ia}\otimes g_{ja}\mid MS\rangle$~~~&~~~* $\ord{1}  {\mathbf {1 }}+\ord{1/N_c^2}$ ~~~~ & ~~~$\ord{1}$ \\
\hline 
\end{tabular}}
\end{table}
From Eqn.~(18) and  Table~1  the mixings involving the GS are as follows:  the $\Delta\ell=0,~{\rm and}~2$ mixings with MS states  are  $\ord{1/\sqrt{\nc}}$, and the $\Delta\ell=2$ mixings with S states are $\ord{1/{\nc^{\frac{3}{2}}}}$. As expected, these mixings can affect the GS masses at best at $\ord{1/\nc}$. One interesting consequence of the relevance of mixing is that it provides the dominant contribution to the electric quadrupole moment of GS baryons (of $\Delta$ for $\nc=3$). 
In the valence quark picture the quadrupole moment operator is given by:
\beq
Q_{ij}(x)=(3\;x_i\,x_j-x^2\; \delta_{ij})\;\hat{Q},
\eeq
where $\hat{Q}$ is the matrix of quark charges. It is necessary to have $\Delta \ell=2$ mixing in order to
obtain non-vanishing quadrupole moments of the GS baryons. The dominant  such mixing just found is the one with MS states.  If one does not re-scale the electric charges with $\nc$, the matrix elements of the charge, namely $\langle MS\mid \hat{Q} \mid S\rangle$  (defined in analogous manner as the 1-body matrix element in (12)), are $\ord{1}$. It is then straightforward to show that the matrix elements of  $Q$ in the physical GS baryons is $\ord{1}$. If one re-scales the charges in the standard fashion one arrives at the known results in Refs.~\cite{Quadrupole}.

The configuration mixings between excited states driven by  2-body operators are determined by the following matrix elements:
\beqa
\langle \Psi'' \mid H^{\rm mix}_{\rm 2-body}\mid \Psi'\rangle &=&\int d^3x\; d^3 y \;(\phi^*(x) \phi''^*(y)+x\leftrightarrow y)\nonumber\\
&\times& L_{ij}(x,y)\; \phi(x) \phi'(y)\; \langle \chi''\mid {\mcal{S}}_{ij} \mid \chi' \rangle.
\eeqa
Using here the  results displayed in  Table 1, one finds that $\Delta \ell=0$ mixings, which require  S-MS mixing,  are $\ord{1/\nc}$  if $\ell=0$ and $\ord{1}$ if $\ell>0$, while $\Delta \ell=2$ mixings 
of type S-S are $\ord{1/\nc^2}$, of type MS-MS are $\ord{1}$ and of type S-MS are $\ord{1/\nc}$. 
These  $1/\nc$ counting results  for  mixings  are  summarized in Table 2.
\begin{table}[h]
\caption{Summary of $1/\nc$ power counting for  configuration mixings. }
{\setlength{\tabcolsep}{0.6em}
\begin{tabular}{@{}ccc@{\hspace{0.12em}}c@{\hspace{0.12em}}l@{}}
\hline & &\\
& S & MS\\  
& & \\ \hline
GS & $\Delta\ell=2:~~\ord{\nc^{-\frac 32}}$& $\ord{\nc^{-\frac 1 2}}$\\ &&\\
& &  $\ell=0,~\Delta\ell=0:~\ord{1/\nc}$\\
S &  ~~~~ $\Delta\ell=2:~\ord{1/\nc^{2}}$~~~~&   $\ell\neq 0,~\Delta\ell=0:~\ord{1}$  \\
&&  $\Delta\ell=2:~~\ord{1/\nc}$\\ &&\\
MS &- &    $\Delta\ell=2:~\ord{1}$  \\
\hline 
\end{tabular}}
\end{table}

An important fact  shown by this analysis is that the $\ord{1}$  mixings between excited states {\it always} involve the coupling to orbital degrees of freedom. If these mixings would have natural size, one would expect that excited baryons would not show the striking pattern of states that can be accommodated into multiplets of $O(3)\times SU(6)$. This is indicating  that the $\ord{1}$  mixings
are small and is in line with the observation  from  baryon masses  that  spin-orbit couplings are dynamically  suppressed. 

\section{Conclusions}
The applications of the $1/\nc$ expansion to baryons have shown the viability of this approach in the real world with $\nc=3$. Quite in general, applications to ground state and excited baryons alike show
 no violations of the hierarchical order implied by the expansion. For instance, applications to masses and decays show that the effective constants obtained by fitting to data never violate the naturalness implied by that hierarchy, {\it i.e.}, these constants are never anomalously large. There are however
 dynamical QCD effects that {\it suppress} some effective constants, the most notable of them being the constants that determine the spin-orbit type couplings. In this sense, the $1/\nc$ expansion serves as a very useful tool to  identify such dynamical suppressions. 
 
 The results reported here addressed the $1/\nc$ expansion in decays and configuration mixings. For the former, we have shown the dominance of one-meson emission in decays of excited baryons, a pattern that  seems to be realized  according  to the  somewhat limited  experimental information on partial widths. For the latter, we have classified the possible configuration mixings and established among other things that all zeroth order  mixings involve spin-orbit type couplings. The striking arrangement of the known excited baryons into multiplets of $O(3)\times SU(6)$  as  shown by the  analyses of masses and decays, implies  that such zeroth order mixings are suppressed.

\section{Aknowledgements}
I  have benefitted from discussions   with Richard Lebed, Carlos Schat and Norberto Scoccola. I especially thank the ECT* for providing the venue for a  stimulating meeting and for partial financial support. This work was  supported by the National Science Foundation  through grant PHY-0300185 and  by  the Department of Energy  contract DE-AC05-84ER40150  under which SURA operates the Thomas Jefferson National Accelerator Facility


\begin{thebibliography}{99}
 \bibitem{tHooft} G.~'t Hooft, {\it Nucl.~Phys.} B {\bf 72}, 461 (1974).
\bibitem{ChPTtalks} See contributions to these proceedings by R.~Kaiser, S.~Peris, and J.~Prades et al.,  and references therein.
\bibitem{Witten} E.~Witten, {\it Nucl.~Phys.} B { \bf 160}, 57 (1979).
 \bibitem{JLG2} J.~L.~Goity,  {\it Yad.~Fiz.} {\bf 68}, 655 (2005)  [hep-ph/0405304].
\bibitem{ConsistencyRel}
J.~L.~Gervais and B.~Sakita,
{\it Phys.~Rev.~Lett.} { \bf 52}, 87 (1984);
{\it Phys.~Rev.} D  { \bf  30}, 1795 (1984).\\
R.~Dashen and A.~V.~Manohar,
{\it Phys.~Lett.} B  { \bf  315}, 425 (1993);
{\it Phys.~Lett.} B { \bf  315}, 438 (1993).\\
A.~V.~Manohar, these proceedings and references therein.
\bibitem{Masses} J.~L.~Goity, 
{\it Phys.~Lett.} B  { \bf  414}, 140 (1997).\\
C.~E.~Carlson, C.~D.~Carone, J.~L.~Goity and R.~F.~Lebed,
{\it Phys.~Lett.} B { \bf  438}, 327 (1998);
{\it Phys.~Rev.} D { \bf  59}, 114008 (1999).\\
C.~L.~Schat, J.~L.~Goity and N.~N.~Scoccola,
 {\it Phys.~Rev.~Lett.}
{\bf  88}, 102002 (2002); 
J.~L.~Goity, C.~L.~Schat and N.~N.~Scoccola,  
{\it Phys.~Rev.} D {\bf  71}, 034016 (2005).
{\it Phys.~Rev.} D {  \bf  66}, 114014 (2002); J.~L.~Goity, C.~L.~Schat and N.~N.~Scoccola,  
{\it Phys.~Lett.} B  {\bf 564}, 83 (2003).\\
N. Matagne and Fl. Stancu, 
{\it Phys.~Rev.} D  {\bf  71}, 014010 (2005), and these proceedings.
 \bibitem{ChQM} A.~V.~Manohar and H.~Georgi, 
 {\it Nucl.~Phys.} B  { \bf 234}, 189 (1984).
  \bibitem{Lebed} R.~F.~Lebed, these proceedings and references therein.
 \bibitem{CaroneDecays}C.~D.~Carone, H.~Georgi, L.~Kaplan and D.~Morin, 
 {\it Phys.~Rev.} D { \bf  50}, 5793 (1994).
\bibitem{GSS4} J.~L.~Goity, C.~L.~Schat and N.~N.~Scoccola,  
 {\it Phys.~Rev.} D {\bf 71}, 034016 (2005).\\
 N.~N.~Scoccola, these proceedings.
 \bibitem{CarlsonCarone} C.~E.~Carlson and C.~D.~Carone, {\it Phys.~Lett.} B  {\bf  484}, 260 (2000).
 \bibitem{SGDecays} N.~N.~Scoccola and J.~L.~Goity, in preparation.
\bibitem{PDG} Particle Data Group, S. Eidelman et al., 
{\it Phys.~Lett.} B {\bf 592}, 1 (2004).
\bibitem{PirjolYan} D.~Pirjol and T.~M.~Yan,
{\it Phys.~Rev.} D { \bf 57}, 1449 (1998);
{\it Phys.~Rev.} D  { \bf 57}, 5434 (1998).
 \bibitem{PirjolSchat} D.~Pirjol and C.~L.~Schat, 
 {\it Phys.~Rev.} D  {  \bf 67}, 096009 (2003).
\bibitem{CohenLebedNelore} T.~D.~Cohen, D.~C.~Dakin, A.~Nellore and R.~F.~Lebed, 
{\it Phys.~Rev.} D  { \bf 69}, 056001 (2004).
\bibitem{Quadrupole} A.~Buchmann and R.~F.~Lebed, 
 {\it Phys.~Rev.} D  { \bf  62}, 096005 (2000).\\
E.~Jenkins, X-d.~Ji and A.~V.~Manohar, 
{\it Phys.~Rev.~Lett.} {\bf  89}, 242001 (2002).
\end{thebibliography}
\end{document}